\documentclass{Interspeech}
\interspeechcameraready 
\title{Replay Attacks Against Audio Deepfake Detection}
\author[affiliation={1,3}]{Nicolas}{Müller}
\author[affiliation={2,3}]{Piotr}{Kawa}
\author[affiliation={1}]{Wei-Herng}{Choong}
\author[affiliation={4}]{Adriana}{Stan}
\author[affiliation={3}]{Aditya}{Tirumala Bukkapatnam}
\author[affiliation={5,6}]{Karla}{Pizzi}
\author[affiliation={1}]{Alexander}{Wagner}
\author[affiliation={1}]{Philip}{Sperl}

\affiliation{}{Fraunhofer AISEC, Germany
  $^2$Wrocław University of Science and Technology, Poland\\
  $^3$Resemble AI, USA
  $^4$Technical University of Cluj-Napoca, Romania\\
  $^5$Neodyme AG, Germany
  $^6$TU Munich, Germany}{}

\email{nicolas.mueller@aisec.fraunhofer.de}

\usepackage{placeins}

\usepackage{comment}
\usepackage{multirow}
\usepackage[table]{xcolor}

\usepackage{booktabs}
\usepackage{algorithm}
\usepackage{algpseudocode}
\usepackage{cleveref}
\usepackage{url}
\usepackage{todonotes}
\PassOptionsToPackage{hyphens}{url}\usepackage{hyperref}
\usepackage{breakcites}\usepackage{xcolor}
\usepackage{cite}

\begin{document}

\maketitle

\begin{abstract}
We show how replay attacks undermine audio deepfake detection: 
By playing and re-recording deepfake audio through various speakers and microphones, we make spoofed samples appear authentic to the detection model. 

To study this phenomenon in more detail, we introduce \emph{ReplayDF}, a dataset of recordings derived from M-AILABS and MLAAD, featuring 109speaker-microphone combinations across six languages and four TTS models. 
It includes diverse acoustic conditions, some highly challenging for detection.

Our analysis of six open-source detection models across five datasets reveals significant vulnerability, with the top-performing W2V2-AASIST model's Equal Error Rate (EER) surging from $4.7\%$ to $18.2\%$.
Even with adaptive Room Impulse Response (RIR) retraining, performance remains compromised with an $11.0\%$ EER.
We release \emph{ReplayDF} for non-commercial research use.
\end{abstract}

\section{Introduction}

Text-to-speech (TTS) and Voice Conversion (VC) technologies have facilitated numerous advancements across various domains.
In the entertainment industry, they enable independent studios in television and film to generate a diverse array of character voices, including the recreation of legacy character voices.
In the healthcare sector, TTS has demonstrated significant potential in assisting individuals with speech impairments, as illustrated by initiatives such as Google’s Parrotron~\cite{parrotron}.
Despite its potential, speech synthesis also presents significant risks, particularly through the misuse of replicating an unknowing target speaker's identity, also known as deepfake technology. These artificially generated identities can be exploited to spread disinformation, undermine trust in authorities, and harm individuals' reputations~\cite{deep_misinfo,df_taylor}. 
Deepfakes have also been employed in fraudulent schemes, voice phishing attacks (vishing), and even in state-on-state warfare~\cite{deepfake_fraud,deepfake_stock,zel_fake}. 
Addressing these challenges requires robust deepfake detection systems to determine whether an audio sample is genuine (\emph{bona~fide}) or fabricated (\emph{spoofed}). 

In the domain of speaker identification and authentication, three distinct attack scenarios emerge: a) \textit{physical attacks}: used for access spoofing, this involves replay attacks targeting voice-biometric systems through the playback of recorded bona~fide utterances; b) 
\textit{logical access}: TTS and VC technologies are used to circumvent voice biometry systems by synthesizing a target speaker's voice characteristics; and c) \textit{deepfake audio}: synthetic voice content designed to deceive human listeners~\cite{asvspoof-2021}, distributed primarily through social media platforms.

\begin{figure}[t]
    \centering
    \input{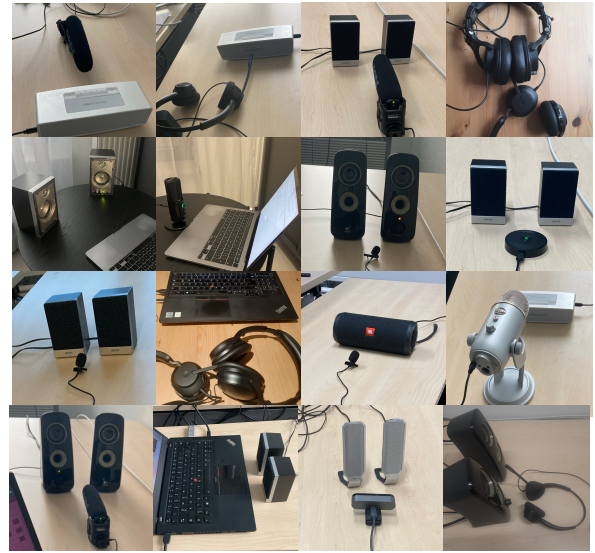}
    \caption{Photographs from the recording labs, where we play bona~fide and spoofed audio samples over a loudspeaker, and record them via microphone. 
    }
    \label{fig:image_recordings}
\end{figure}

In this context, we identify a critical gap: the vulnerability of deepfake detection systems to physical or replay attacks. 
Specifically, we examine scenarios where attackers attempt to compromise the  detection system by re-recording the synthesised audio. 
Our analysis demonstrates that such replay attacks can successfully disguise audio deepfakes as genuine recordings, likely by removing subtle artifacts that detection models rely on for identification.

To address this research gap, 
we:

\begin{itemize}
    \item introduce \emph{ReplayDF}, a comprehensive dataset\footnote{\url{https://deepfake-total.com/replay_df}} of deepfake recordings. It comprises 109combinations of loudspeaker and microphone across six languages and four text-to-speech (TTS) systems, plus Mean Opinion Scores;
    \item evaluate existing audio deepfake detection models using this dataset, demonstrating that \emph{replay attacks} effectively disguise fake audio as genuine;
    \item show a correlation between model performance and recording quality; and
    \item provide evidence that adaptive retraining using room impulse responses (RIRs) from \emph{ReplayDF} can help to lessen the effects of replay attacks.
\end{itemize}

\FloatBarrier

\section{Related Work}

Audio deepfake detection has been largely driven by ASVspoof~\cite{asvspoof-2021,asvspoof-2019,asvpaper24}, a challenge that initially focused on voice biometrics but recently expanded to include deepfake detection. The urgency of the problem is underscored by the rise of commercial text-to-speech providers like Resemble AI, Respeecher, and ElevenLabs~\cite{resemble-ai,respeecher,eleven-labs}, some of which have been involved in misinformation campaigns \cite{Research6:online}.
Research advancements in neural audio deepfake detection span both front-end and back-end components. Front-ends extract features from audio, evolving from time-frequency representations like mel-spectrograms~\cite{lcnn} to raw waveforms (such as RawNet2~\cite{rawnet-2}) and self-supervised learning methods like Wav2Vec2~\cite{baevski2020wav2vec}. Back-ends focus on classifying the samples suing the extracted features, and exploring advanced neural architectures such as graph attention networks and conformers~\cite{aasist,truong2024temporal}.
Challenges in audio deepfake detection include generalization, which evaluates the discrepancy in performance on seen versus unseen data~\cite{muller2022interspeech, pascu_is24}, explainability of the model's decision~\cite{muller2021speech,channing2024toward}, as well as vulnerability to replay attacks.

Replay attacks were originally formulated in the ASVspoof challenge~\cite{asvspoof-2019} in the context of ``liveliness detection'' to study the vulnerability of speaker authentication systems against recorded voice trials.
The concept of recorded audio is also relevant in the context of adversarial evasion attacks, where the goal is to create an imperceptible perturbation $\delta$ such that a given model $f$ misclassifies an input $x$, i.e. $f(x + \delta) \not = y$ such that $\delta$ is small w.r.t some norm.
It has been shown that the creation of adversarial samples that survive the ``air-gap'' is especially challenging and requires specialized techniques~\cite{liang2022adversarial,qin2019imperceptible,athalye2018synthesizing}. 
This is because when presenting and capturing such an adversarial sample, the perturbation $\delta$ can be lost or altered in the process, leading to the deliberate manipulations of the original input to no longer be effective.
In the context of replay attacks on audio deepfake detection, there is little related work.
Luong et al.~\cite{luong2024room} demonstrate that RIRs degrade the performance of detection systems on ASVspoof 2021. However, their study is limited to simulated recordings rather than real-world replays.

This work investigates the replay attack phenomenon into greater depth by conducting controlled ``air-gapping'' experiments in a laboratory environment, where audio deepfakes are played back and re-recorded. 


\section{ReplayDF Database}

To systematically assess the impact of replay attacks on deepfake detection, we introduce \emph{ReplayDF}, a dataset of audio recordings generated by playing and re-recording both bona~fide and spoofed samples using a diverse set of loudspeakers and microphones. 
The dataset spans six languages and incorporates attacks from four TTS models, ensuring an equal distribution between spoof and bona~fide samples.

The data generation pipeline, outlined in \Cref{alg:tts-experiment}, systematically selects audio samples from the MLAAD v5 dataset for spoofed recordings and the M-AILABS dataset for bona fide samples. Each selected sample is played through a loudspeaker and recorded using a microphone, thereby capturing playback distortions introduced by real-world acoustic environments and hardware characteristics.
This procedure is repeated across multiple recording setups, resulting in a dataset encompassing 109 unique recording configurations and totaling 132.5 hours of audio data.
Each audio sample is accompanied by complete metadata, as listed in \Cref{tab:attributes_values}: the original and recorded file paths, attack type (i.e., the TTS model if the original audio is spoofed), language, recording hardware, and setup images. 
A combination of loudspeaker and microphone is referred to as a \emph{setup}, uniquely identified by a hash-based \emph{uid}. 
To evaluate the quality of the dataset, we perform subjective listening tests and rate the quality of the recordings using Mean Opinion Scores (MOS), as detailed in \Cref{eval:mos}.

\begin{table}[t]
    \centering
    \begin{tabular}{lp{0.35\textwidth}}
    \toprule
    \textbf{Attribute} & \textbf{Info} \\ 
    \midrule
    original\_file & Relative path to original file from either \texttt{MLAAD} or \texttt{M-AILABS}. \\ 
    recorded\_file & Relative path to recorded file. \\ 
    label & \texttt{spoof} or \texttt{bona~fide}. \\ 
    architecture & \texttt{bark}, \texttt{vits}, \texttt{XTTS v1.1}, or \texttt{XTTS v2.0}. \\ 
    language & \texttt{en}, \texttt{de}, \texttt{fr}, \texttt{it}, \texttt{pl}, or \texttt{sp}. \\ 
    mic & Description of the microphone used. \\ 
    speaker & Description of the loudspeaker used. \\ 
    uid & Unique folder identifier. \\ 
    setup.jpg & Photograph of the recording setup. \\ 
    RIR.wav & Room Impulse Response of recording setup. \\
    \bottomrule
    \end{tabular}%
    \caption{Metadata for \emph{ReplayDF}.}
    \label{tab:attributes_values}
\end{table}

\begin{algorithm}[t]
\label{alg:tts-experiment}
\begin{algorithmic}[1]
\State $I = [id_0, id_1, ..., id_{108}]$  \Comment{Speaker/microphone used.}
\State $n = 10$ 
\State $R = \{ \}$ \Comment{List to hold recorded audio files.}
\State $O = \{ \}$ \Comment{List to hold original audio files.}
\For{id $\in$ I}
    \For{lang $\in$ \{en, de, fr, it, pl, es\}}
        \For{model $\in$ \{bark, vits, xtts\_v1.1, xtts\_v2.0\}}
            \State $A \gets$ choose $n$ from \texttt{M-AILABS(lang)}
            \State $B \gets$ choose $n$ from \texttt{MLAAD(lang, model)}
            \For{w $\in$ $A \cup B$}
                \State r = \texttt{play\_and\_record(id, w)}
                \State $R \gets R \cup \{r\}$
                \State $O \gets O \cup \{s\}$
            \EndFor
        \EndFor
    \EndFor \Comment{$6\cdot 4\cdot 2\cdot 10 = 480$ recordings per $id$.}
\EndFor \Comment{Total number of recordings: $|I| \cdot 480 = 52320$.} 
\end{algorithmic}
\caption{The data creation pipeline for \emph{ReplayDF}. 
For each setup (i.e., per combination of loudspeaker and microphone), we select $n=10$ instances for each language and TTS model from both MLAAD v5 (spoof) and M-AILABS (bona~fide).
Recordings and original audio files are stored to create a balanced dataset of air-gapped vs. non-air-gapped data.
}
\end{algorithm}

\section{Experiments}
\subsection{Evaluation Approach}

We evaluate \emph{ReplayDF} across multiple scenarios to assess the impact of replay attacks on audio deepfake detection models.
We define two key data partitions:
first, \emph{ReplayDF} (set $R$): All audio files generated as in \Cref{alg:tts-experiment}, containing equal amounts of bona~fide and spoofed instances.
Second, the \emph{Baseline} dataset (set $O$): the original input instances from MLAAD v5 and M-AILABS, serving as a comparative baseline against \emph{ReplayDF}.

\subsection{Publicly Available Deepfake Detection Checkpoints}

The first evaluation we perform with \emph{ReplayDF} is against open-source deepfake detection models. We use publicly available checkpoints and their original hyperparameters and rate their performace for the $O$ and $R$ subsets of \emph{ReplayDF}.
\Cref{tab:opensource} summarizes the results, highlighting performance differences in percentage points between the two subsets.
Across all tested models, we observe a significant performance degradation of up to $20$ percentage points when transitioning from \emph{Baseline} to \emph{ReplayDF}.

\begin{table}[t]
    \centering
    \resizebox{.475\textwidth}{!}{

\begin{tabular}{lcccc}
\toprule
\multirow{2}{*}{\textbf{Model}} & \multicolumn{2}{c}{\textbf{Accuracy (\%)} $\uparrow$} & \multicolumn{2}{c}{\textbf{EER (\%)} $\downarrow$} \\
\cmidrule(lr){2-3} \cmidrule(lr){4-5}
 & Baseline & ReplayDF & Baseline & ReplayDF \\
\midrule
Whisper~\cite{whisper-df}           &  57.9 & 50.0 & 44.7 & 49.5 \\
Raw PC Darts~\cite{raw-pc-darts}    &  69.4 & 56.6 & 32.1 & 43.9 \\
RawNet2~\cite{rawnet-2}             &  74.3 & 57.1 & 25.9 & 43.1 \\
TCM ADD~\cite{truong2024temporal}   &  73.5 & 59.6 & 13.3 & 37.3 \\
RawGAT-ST~\cite{rawgat-st}          &  79.4 & 58.7 & 19.8 & 40.2  \\
W2V2-AASIST~\cite{tak2022automatic} &  \cellcolor{blue!25}90.0 & 74.2 & \cellcolor{blue!25}10.6 & 24.8 \\
\bottomrule
\end{tabular}

    }
    \caption{Performance of Open-Source models with publicly available checkpoints in mean accuracy and EER over ReplayDF, as well as the original audio files (Baseline). The threshold for accuracy computation is as specified in the respective original publication. 
    In all scenarios, replay attacks deteriorate model performance.
    }
    \label{tab:opensource}
\end{table}

\subsection{Dataset Selection}

\begin{table}[t]
\centering
\resizebox{.475\textwidth}{!}{
\begin{tabular}{lcccc}
\toprule
\multirow{2}{*}{\shortstack{\textbf{Training} \\ \textbf{Dataset}}} & \multicolumn{2}{c}{\textbf{Accuracy (\%)} $\uparrow$} & \multicolumn{2}{c}{\textbf{EER (\%)} $\downarrow$} \\
\cmidrule(lr){2-3} \cmidrule(lr){4-5}
 & Baseline & ReplayDF & Baseline & ReplayDF \\
\midrule
ASVspoof 19 & 74.6 ± 6.6 & 55.1 ± 3.2 & 14.3 ± 7.7 & 34.7 ± 3.1 \\
ASVspoof 5  & 67.6 ± 8.6 & 52.8 ± 1.4 & 12.5 ± 4.1 & 34.8 ± 3.5 \\
Fake-or-Real   & 54.2 ± 0.9 & 49.5 ± 0.0 & 28.4 ± 1.6 & 45.0 ± 1.3 \\
In-the-Wild   & 66.8 ± 3.1 & 52.9 ± 1.4 & 30.2 ± 3.2 & 42.4 ± 1.2 \\
ODSS  & \cellcolor{blue!25} 94.7 ± 0.7 & 77.7 ± 2.4 & \cellcolor{blue!25} 4.7 ± 0.9 & 18.2 ± 1.5 \\
\bottomrule
\end{tabular}
}
\caption{Performance of W2V2-AASIST, trained on five different datasets, and evaluated on \emph{ReplayDF}. Results computed over three independent trials, with mean and standard deviation shown.}
\label{tab:six_models}
\end{table}

\begin{table}[t]
    \centering
    \resizebox{.475\textwidth}{!}{

\begin{tabular}{lrrrrr}
\toprule
\multirow{2}{*}{\shortstack{\textbf{Attack}}} & \multicolumn{2}{c}{\textbf{No Augmentation}} & \multicolumn{2}{c}{\textbf{With Augmentation}} \\
\cmidrule(lr){2-3} \cmidrule(lr){4-5}
 & Baseline & ReplayDF & Baseline & ReplayDF \\
\midrule
Bark & 82.6 & 40.7 & 83.0 & 56.9 \\
VITS & 82.2 & 53.4 & 75.7 & 65.8 \\
XTTS v1.1 & 100.0 & 66.3 & 99.7 & 76.2 \\
XTTS v2 & 100.0 & 59.4 & 99.8 & 73.6 \\ \midrule
bona fide & 98.3 & 97.7 & 99.9 & 98.6 \\
\bottomrule
\end{tabular}

    }
    \caption{
    Detection accuracy [\%] ($\uparrow$) of W2V2-AASIST on the \emph{Baseline} and \emph{ReplayDF}, with and without RIR augmentation. While TTS-generated spoofs cause substantial accuracy reductions (up to $-43.6$ percentage points), bona~fide detection remains unaffected. 
    Incorporating RIR augmentation during training reduces the effect of the replay attacks.
    }
    \label{tab:res_odss}
\end{table}

To evaluate this phenomenon in more detail, we chose the best-performing model architecture from \Cref{tab:opensource}, i.e. W2V2-AASIST,  and re-train it on five different audio deepfake datasets:
ASVspoof2019~\cite{asvspoof-2019}, ASVspoof 5~\cite{asvpaper24}, Fake-or-Real~\cite{for}, In-the-Wild~\cite{muller2022interspeech} and the Open Dataset of Synthetic Speech (ODSS)~\cite{odss}.
The models are trained for $75$ epochs using early stopping based on the training loss, employing the Adam optimizer with a learning rate of $4 \cdot 10^{-6}$ and a batch size of $32$.

We then evaluate the retrained models on \emph{ReplayDF} and the \emph{Baseline} dataset.
\Cref{tab:six_models} presents the results in terms of mean and standard deviation of the accuracy and EER, over three independent trials.
Given that the ODSS-trained model performs best, we select this configuration for the following evaluations.
Note that this experiment re-confirms the results from \Cref{tab:opensource}:
irrespective of training dataset, the model performance deteriorates when moving from the \emph{Baseline} dataset to \emph{ReplayDF}. Even in the case of the best model trained on ODSS, the deterioration is significant, dropping from $4.7\%$ to $18.2\%$ EER.

\subsection{Analysis of False Positives and Negatives}

An essential analysis is to evaluate whether the overall decrease in model performance is caused by an increased number of false positives or false negatives. In other words: are bona~fide instances mistakenly classified as spoof (false positives) or are the spoofed instances missed by the model (false negatives)?
\Cref{tab:res_odss} details the results obtained by W2V2-AASIST model trained on ODSS.
It shows the model accuracy\footnote{Since we compute the performance per attack, we only have a single label and cannot compute EER. 
Instead, we report accuracy, where predictions are classified as positive if they exceed a threshold of 50\%.
Notably, since \emph{ReplayDF} is perfectly balanced, accuracy serves as an appropriate metric in this context.} 
for each of the individual attacks in \emph{ReplayDF} (Bark, VITS, XTTS v1.1 and v2.0) and for the bona~fide instances.
It can be clearly observed that the air-gap affects solely the spoofed instances, which exhibit a decrease in performance between $28$ and $42$ absolute percent points, while the performance over the bona~fide instances remains unchanged.

Finally, we note that models trained on datasets where noise patterns differ between genuine and spoofed audio (like older datasets such as In-the-Wild or ASVspoof 19) can develop problematic shortcuts~\cite{geirhos2020shortcut,muller2021speech}: they learn to equate poor audio quality with spoofed samples. This leads to a misclassification of the entire \emph{ReplayDF} data as spoofed audio, simply because recordings (both bona~fide and spoof) contain channel noise.

\subsection{Adaptive Defender}
Next, we evaluate if augmenting the training data with room impulse responses (RIRs) from the dataset can help defend against replay attacks.
During training, we convolve the training data with RIRs from \emph{ReplayDF}, then evaluate performance on both the \emph{Baseline} and \emph{ReplayDF} datasets as before.

RIR augmentation reduces the replay attack's impact, improving the overall EER from $18.2\%$ to $11.09\%$ (compared to the baseline EER of $4.7\%$ and $2.2\%$ without and with RIR augmentation, respectively).
The right-hand side of \Cref{tab:res_odss} shows results per attack type.
While \emph{Baseline} performance remains stable, accuracy on \emph{ReplayDF} improves by about $10$ to $15$ percentage points for each of the attacks.
Thus, while RIR augmentation during training does improve resilience to replay attacks, it falls short of fully mitigating this vulnerability.

\begin{figure*}[t]
    \centering
    \includegraphics[width=0.995\linewidth]{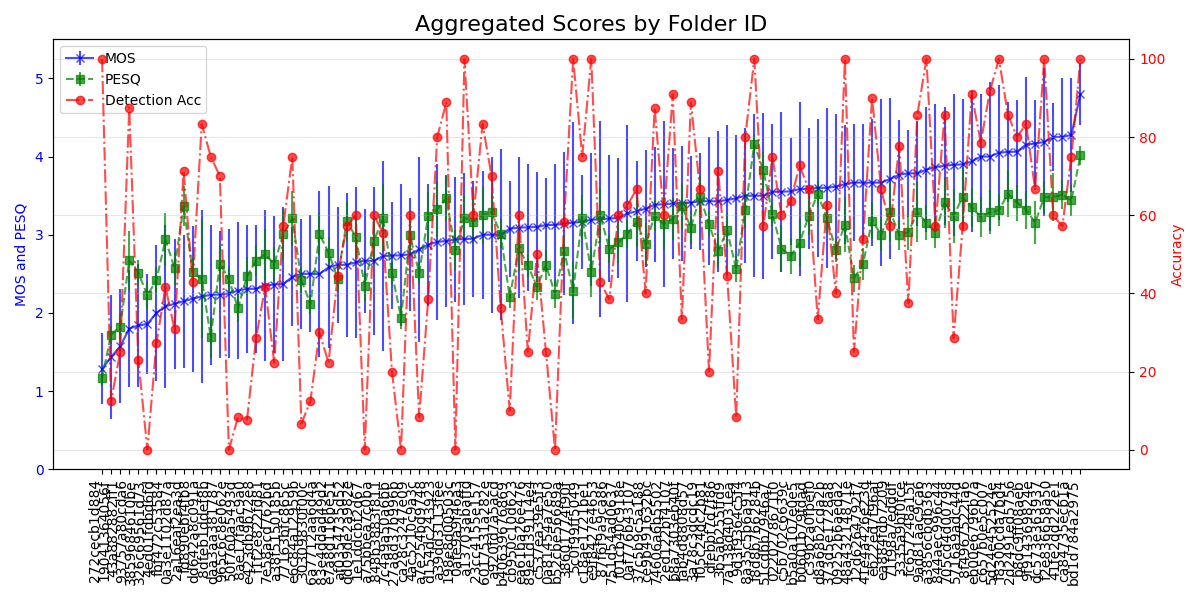}
    \caption{Overview of recording quality in \emph{ReplayDF}, measured by MOS and PESQ (blue, green). Detection performance on audio deepfakes (red) correlates with recording quality, showing Pearson correlations of 0.423 and 0.509, respectively. This suggests that the more aggressive the replay attack, the worse the detection performance.}
    \label{fig:mos_pesq_attack}
\end{figure*}

\subsection{Recording Quality vs. Detection Performance}\label{eval:mos}
To further investigate the models' performance deterioration, we analyze whether it correlates with potential quality degradation in the recordings.
This analysis is conducted using two measures.
First, we use the Perceptual Evaluation of Speech Quality (PESQ)~\cite{rix2001perceptual} score, which is an objective metric used to assess audio quality by comparing a reference and degraded signal.
It provides a score ranging from -0.5 to 4.5, where higher values indicate better quality. 
Second, we perform human listening tests, where subjects are asked to score bona~fide samples from \emph{ReplayDF} on a score between $1$ (worst) and $5$ (best).
We obtain a Mean Opinion Score (MOS) for each setup UID by averaging three scores from at least four listeners per UID.

\Cref{fig:mos_pesq_attack} displays recording UIDs against PESQ and MOS, as well as the accuracy of the W2V2-AASIST deepfake detection system. 
The analysis reveals that our database encompasses a wide range of recording quality levels, with MOS and PESQ values varying from low (i.e. $< 2$) to excellent ($>4$). 
Furthermore, MOS and PESQ scores correlate with a Pearson coefficient of~$0.681$, confirming consistency between subjective and objective quality assessments. 
The deepfake detection accuracy exhibits a Pearson correlation of $0.423$ with MOS and $0.509$ with PESQ.
This suggests that lower-quality replay attacks reduce the detection effectiveness of audio deepfake systems.

\subsection{Recordings vs. Noise}
Finally, we investigate whether the performance drop in deepfake detection is primarily caused by general quality degradation, such as added noise, or rather by the loss of distinctive model- or deepfake-specific characteristics due to the air-gap.
To address this, we add noise to the \emph{Baseline} dataset and evaluate the previously trained model on it. 
We add three different types of noise (Gaussian, white, and pink noise) at varying signal-to-noise ratios (SNR). For each input audio file, a synthetic noise of matching length is generated and mixed with the original audio at a randomly selected target SNR level between 15-40 dB (the SNR value is recorded in the metadata). To achieve the target SNR, we estimate the signal power and generate a noise sample of equal number of samples. Given the noise signal's power and a target SNR value in dB, a linear scaling factor is calculated scale the noise signal down, such that the mixture of the audio and noise signals is at the desired SNR.



We find that model performance is not impacted much. W2V2-AASIST exhibits accuracy changes of $-2.3\%$ (Bark), $-0.2\%$ (XTTS v2), $+0.0\%$ (XTTS v1.1), and $+3.1\%$ (VITS), while bona fide detection remains unaffected ($+0.0\%$).
These findings suggest that the performance degradation observed in replay attack scenarios is not merely due to noise.
Given that Luong et al.~\cite{luong2024room} report similar findings on \emph{simulated} recordings, improving deepfake detection models to better handle convolutional noise---such as distortions introduced by room impulse responses---could be a crucial step toward enhancing their resilience against attacks.


\section{Conclusion}
In this work, we investigate the impact of replay attacks on audio deepfake detection systems by introducing \emph{ReplayDF}, a comprehensive dataset of $132.5$ hours of re-recorded spoof and bona~fide audio. 
Our results demonstrate that replay attacks significantly degrade detection performance, effectively disguising deepfake audio as authentic, while bona~fide samples remain unaffected. 
We further show that simply adding noise does not lead to the same degradation, indicating that replaying removes key artifacts relied upon by detection models. 
To support future research in this area, we publicly release \emph{ReplayDF} along for non-commercial use.
\FloatBarrier
\clearpage
\textbf{Acknowledgement.} 
This work was partially funded by project DLT-AI SECSPP (id: PN-IV-P6-6.3-SOL-2024-2-0312) and the Department of Artificial Intelligence, Wrocław University of Science and Technology.

\bibliographystyle{IEEEtran}
\bibliography{references}

\end{document}